\documentstyle[preprint,aps]{revtex}

\begin{document}
\draft

\title{
KINETIC PROPERTIES OF A BOSE-EINSTEIN GAS AT FINITE TEMPERATURE
}

\author{Teresa Lopez-Arias and  Augusto Smerzi} 

\address{International 
School of Advanced Studies, Via Beirut 2/4, I-34014, Trieste Italy } 

\date{\today}
\maketitle

\begin{abstract}
We study, in the framework of the Boltzmann-Nordheim equation (BNE),
the kinetic properties of a boson gas above the Bose-Einstein 
transition temperature $T_c$.
The BNE is solved numerically within a new algorithm, that has been tested with
exact analytical results for the
collision rate of an homogeneous system in thermal equilibrium.
In the classical regime ($T > 6~ T_c$), 
the relaxation time of a quadrupolar deformation in momentum space 
is proportional to  
the mean free collision time $\tau_{relax} \sim T^{-1/2}$.
Approaching the critical temperature ($T_c < T < 2.7~ T_c$), 
quantum statistic effects in BNE become dominant, and 
the collision rate increases dramatically.
Nevertheless, 
this does not affect the relaxation properties 
of the gas that depend only on the spontaneous 
collision term in BNE. The relaxation time 
$\tau_{relax}$ is proportional to $(T - T_c)^{-1/2}$, 
exhibiting a critical slowing down.
These phenomena can be experimentally confirmed looking at
the
damping properties of collective motions induced on
trapped atoms. The possibility to observe a transition  
from collisionless (zero-sound) to hydrodynamic (first-sound)  is
finally discussed.
 
\end{abstract}
\pagebreak
\narrowtext

The study of the kinetic properties of a (non-condensate)
Bose-Einstein (B-E) gas, being interesting in itself, is becoming 
particularly 
important in relation with the recent experimental observations of
the Bose-Einstein condensation (BEC) of 
trapped atoms \cite{1,2,3}
and excitons \cite{4}. 
In the experimental setups of \cite{1,2,3}, a gas of alkali atoms 
is trapped and cooled down below the BEC critical temperature 
($ T \simeq 100~ nK$ for $N \simeq 10^4$ atoms in \cite{1}). 
The cooling is achieved tuning
the trapping magnetic field in order to allow  
the most energetic particles to escape, with  
relaxation processes(essentially one- and 
two-body collisions) 
driving the system to 
new equilibrium states at lower temperatures. 
It is clear that the understanding of the dynamics of the
condensation, requires 
a kinetic model describing the interplay between the time scale associated
with the evaporative cooling, the relaxation processes
and the emerging of the condensate
\cite{5,6,7,8}. Such a complete analysis is still lacking \cite{6}.

One possible scenario is to assume that the time scales 
associated with the growth of
the condensate
are large compared with the relaxation to equilibrium. 
The condensation will have a quasi-static character 
allowing the description of the system in the framework of the 
grand-canonical statistical 
ensemble \cite{9,10,11}. 
On the other hand, for such dilute systems 
(the mean inter-atomic distance is much greater then the scattering length), 
the equilibration times can be far longer than the time 
associated with the growth of the condensate. In this case,
non-equilibrium processes would be crucial in understanding the
BEC. Yet, the 
finite life-time of the trapped atoms is in competion with
the time necessary for the emergence of the condensate.
This particular problem explains the so far unfruitful search 
of condensation in a trapped gas of Hydrogen atoms \cite{12}.  
A further intriguing possibility, is the observation of the
transition from collisionless (zero-sound) to hydrodynamic 
(first-sound) modes, that occurs when the number of collisions is
enough to reach 
local thermodynamic equilibrium.

Several theoretical papers address the dynamics of a weakly 
interacting B-E gas. Levich and Yakhot consider its evolution 
in the presence of 
a thermal bath \cite{13},
finding that the time for condensation is infinite. 
Snoke and Wolfe \cite{14} solve numerically the BNE 
for an homogeneous, thermally isolated, boson gas.
They demonstrate that in the classical
regime (high temperatures and/or low densities) the Maxwell-Boltzmann
distribution 
is achieved within few inter-atom scattering times.
At sufficiently high densities, in the degenerate regime,
the collision rate 
increases, but  
the number of collisions
necessary to relax the system 
increases as well. The net
result is that the relaxation time remains proportional to the 
classical mean collision time.
Again, the appearance of a B-E condensate (an infinite value of the
distribution at zero momentum) is not achieved.
However, a different conclusion is reached by
Semikoz
and Tkachev \cite{15}, who 
found the signature of the B-E condensation in a finite time.

The authors of \cite{14,15} consider  
an isotropic deformation in momentum space. In this case
it is possible to simplify greatly the collision integral in 
BNE, making feasible the numerical calculations in a grid of discrete
energy points. 

In this paper we study, as a function of the temperature,
 both the collision rate of a B-E gas at equilibrium
and the relaxation time of a quadrupolar deformation
in the momentum space.
Differently from the cases 
studied in \cite{14,15} we need to solve the BNE in the full momentum space,
without assuming the isotropy of the system. To this end, we develop
a new numerical algorithm, that we test with analytical results. 
Our analsys indicates that for a quadrupolar (non-isotropic)
deformation the relaxation times diverges as 
$\tau_{relax} \propto (T - T_c)^{-1/2}$, exhibiting a critical slowing down.
This could seem in contradiction with the dramatic increase of
the 
collision rate that occurs approaching 
the critical temperature.
However, the effect is easily understood
as due to the cancellations between
the stimulated "gain" and "loss" collision terms in the BNE, 
with the relaxation time that depends only on the "classical"
spontaneous collision term.
As a consequence, close to the critical temperature $T < 2.7~T_c$
the system exhibits 
zero-sound modes, and the simple criteria given in the literature that
state the hidrodynamic limit occurs when $\omega \tau_{coll} <<1$, with $\omega$
the frequency of the collective motions, clearly breaks down.  

We note that the BNE approach cannot describe 
the onset of the B-E phase transition. In fact, the emerging of 
coherence (with all atoms in the condensate having a common, 
well defined phase) 
requires higher order correlations \cite{7} not included in BNE, 
that is derived
in the random phase (semiclassical) approximation \cite{6}.

The experimental study of kinetic properties of the B-E
gas is quite active.
Collective motions of a trapped condensate 
have been induced 
using time-dependent magnetic fields
with different multipolarities \cite{16,17}. 
The frequencies \cite{18,19,20} and the relaxation times 
\cite{19} of these
resonances have been studied
in detail, and the interest is now focusing 
on their finite-temperature properties \cite{21}. 

The modification of the classical Boltzmann equation in order to take 
into account the effect of Bose statistics in
the collision integral was first suggested by Nordheim \cite{22}:

$${\partial{n}\over{\partial{t}}} + { \vec p\over m}
{\partial{n}\over{\partial{\vec r}}}-
{\partial{U}\over{\partial{\vec r}}}~
{\partial{n}\over{\partial{\vec p}}}
= ({\partial{n}\over{\partial{t}}})_{col}$$
$$
 ({\partial{n}\over{\partial{t}}})_{col} = 
 {g \over {(2 \pi \hbar)^6}} \int d {\vec p_2} ~
d {\vec p_3} ~ d {\vec p_4}~ W ~ 
\delta^3(\vec p + \vec p_2 -\vec p_3 -\vec p_4)
~ \delta(\epsilon + \epsilon_2 -\epsilon_3 -\epsilon_4)$$ 
$$n(\vec r, \vec p) n(\vec r, \vec p_2)
[1 + n(\vec r, \vec p_3)] [1 + n(\vec r, \vec p_4)]
-
n(\vec r, \vec p_3) n(\vec r, \vec p_4)
[1 + n(\vec r, \vec p)] [1 + n(\vec r, \vec p_2)]
\eqno(1)$$
with $g$ the spin degeneracy that, later on, will be set $g = 1$. 
The collision rate $W$ is related with the scattering cross-section as 
$$
W = {{8 \pi^2 \sigma \hbar^3}\over m^2}
\eqno(2)
$$

where $\sigma = 8 \pi a^2$, with $a$ the scattering length.
Here we develop a Boson Monte Carlo Dynamics (BMD) method 
to solve this equation for
the homogeneous case (that can 
be generalized to solve the 
BNE eq.(1) in the full phase-space).
We parametrize the one-body distribution function as a sum of 
"weighted-test-particles" (WTP):
$$n(\vec p, t) =  c~ \sum_{i=1}^{N_{TP}}~  w(\vec p- \vec p_i(t))$$
$$c = {\rho_0 \over {N_{TP}}} {(2 \pi \hbar)}^3
\eqno(3)
$$
where $\rho_0$ is the density of the system, and $N_{TP}$ is the total
number of "test-particles".
We choose the weight functions 
$w(\vec p- \vec p_i)= \delta(\vec p- \vec p_i)$ where 
$p_i(t)$ is the time dependent position in momentum space of the 
test-particle $i$.

The time evolution of the particles swarm proceeds along the following steps:

1) the test-particles are randomly distributed in momentum space
 according to the initial distribution 
 $n( \vec p , t=0)$

2) at each time-step $\Delta t$,  
we randomly choose two test-particles $i,j$, and 
we define a collision probability as: 
$$P_{coll} = 1 - exp[ -{{\| \vec p_i - \vec p_j \|} \over m}
~ \rho_0 ~ \sigma ~  F(\vec p^{~'}_i, \vec p^{~'}_j) ~  \Delta t]
\eqno(4)$$ 
where 
$F = [1 + n({\vec p}^{~'}_i)] [1 + n({\vec p}^{~'}_j)]$.

3) A random number $R = \{ 0,1 \}$ is compared with $P_{coll}$: if 
$R \le P_{coll}$ the collision is accepted. 

4) The new momenta ${\vec p}^{~'}_i$ and ${\vec p}^{~'}_j$ 
that the two particles would have if the collision were accepted  
are randomly chosen imposing the conservation of energy and momentum.

4) Repeat the process from 2) until all test particles have had
 the chance to 
collide in that time step.

The total energy, momentum and number 
of particles are exactly conserved
at all time-steps. 
It is clear that $\Delta t~~(\sim 0.01~s)$ and $N_{tp}~~(\sim 1000)$
should be chosen to reach the convergence of the results.
It is especially important for the time step $\Delta t$ to be small
enough in order to ensure 
that the number of two-body collisions (that diverges 
approaching the critical temperature) remains small compared with the total 
number of test-particles ($N_{coll} << N_{TP}/2$) at each time step.
The factor $F$ in eq.(4) gives the stimulated 
collisions induced by the statistics. For fermions 
the minus sign should be used 
instead of the plus sign; $F=1$ leads to  
the classical Boltzmann collision integral.
We note that similar Monte Carlo methods have been
used for fermion
\cite{23,24} and boson \cite{25} systems.

The number of two-body collisions per unit time (collision rate)
as a function of the temperature is given by:
$$
{{d N_{coll}} \over {d \vec r~ dt}} = 
 {1 \over {(2 \pi \hbar)^9}} ~ \int d {\vec p_1} ~  d {\vec p_2}
~ d {\vec p_3} ~  d {\vec p_4} ~ W 
~ \delta^3(\vec p_1 + \vec p_2 -\vec p_3 -\vec p_4)
~ \delta(\epsilon_1 + \epsilon_2 -\epsilon_3 -\epsilon_4)$$ 
$$n(\vec p) n(\vec p_2)
[1 + n(\vec p_3)] [1 + n(\vec p_4)]
\eqno(5)$$
Expanding  the B-E distribution as: $n = (exp[(H - \mu)/T] -1)^{-1} = 
\sum_{j=1}^{\infty} exp(j (\mu - H)/T) $, 
where $H = {p^2 \over {2m}} + U(\vec r)$
with $U$ the external (or mean-field) potential,  
we can carry out analytically all the integral in eq.(5), getting:
$$
{{d N_{coll}} \over {d \vec r dt}} = 
{\sigma \over {(2 \pi \hbar)^6}} {8 \over {\pi m}}
{(2 \pi m T)}^{7/2}
{ \sum_{i,j} {{\sqrt{ i+j}} \over {i^2 j^2}} exp[{ {(i + j) (\mu - U(\vec r))} 
\over T }}]$$
$$+ 2 \sum_{i,j,k} {{\sqrt{ i+j+k}} \over {i j (i+k) (j+k)}}
exp[{ {(i + j+k) (\mu - U(\vec r))} \over T  }]+$$
$$ \sum_{i,j,k,l} {{\sqrt{ i+j+k+l}} \over {(i+k) (j+k)(i+l)(j+l)}}
 exp[{ {(i + j+k+l) (\mu - U(\vec r))} \over T  }]
 \eqno(6)$$

Since we are considering homogeneous systems, we put $U( \vec r) = const.$
In fig.(1) we compare the analytical results
for the normalized collision rate ${dN_{coll} \over {dt}}$ eq.(6)
(dotted line) with the results 
obtained numerically with BMD (solid line). At 
temperatures $T > 6~ T_c$, 
the Bose-Einstein collision rate becomes equal to the  
classical Maxwell-Boltzmann one (dashed line).
At $T = 2.7 ~ T_c$ the B-E collision rate shows a minimum, 
and for $T < 2.7 ~ T_c$ it increases rapidly deviating significantly
from the classical Maxwell-Boltzmann trend. 
The fast increase is due to 
the stimulated collisions induced by the statistical factors
$(1 + n)$ in eq.(5). At high temperatures (classical regime), 
$n << 1$ and these factors
are negligible, while, 
approaching the critical temperature,
$n >>1$
and quantum statistic effects become dominant.
To better illustrate this point, we show in fig.(2) the different
contributions to the collision rate obtained from eq.(5). The long-dashed line
corresponds to the term
coming from $n(\vec p) n(\vec p_2)$. 
This contribution decreases with temperature 
as it does the classical one (Maxwell-Boltzmann) in fig.(1).  
The dotted line corresponds to the term 
$n(\vec p) n(\vec p_2) [n(\vec p_3) + n(\vec p_4)]$, that shows the deviation
from the classical trend. The biggest contribution to the collision rate 
comes from the term 
$n(\vec p) n(\vec p_2) n(\vec p_3) n(\vec p_4)$ (dashed line).

The chemical potential $\mu$ is constrained by the normalization 
condition:
$$\sum_{j=1}^{\infty} exp(j \mu/T) 
j^{(- 3/2)} = ({{2 \pi \hbar^2}\over {m}})^{3 \over 2}
\rho_0 T^{3 \over 2} = 2.6 ({T \over T_c})^{3 \over 2}
\eqno(7)$$
where the critical temperature, obtained setting $\mu = 0$, is 
$T_c = ({2 \pi \hbar^2 \rho_0 / 2.6 ~ m })^{3 \over 2}$. 

Eq.(7) shows that 
${\mu \over T} = G[{T \over T_c}]$ 
where $G$ is a universal function 
independent of the density of the system. This fact allows 
to calculate from fig.(1) the collision rate for an  
homogeneous system with arbitrary density. 
In particular, for a given $T \over {T_c}$, we have:
$$ {{d N_{coll}} \over {d \vec r dt}} \propto \sigma {T_c}^{7/2}
\eqno(9)$$

From the experimental point of view it is of particular importance
to know the relaxation time $\tau_{relax}$, that is, the characteristic time 
for an anisotropic distribution
to reach the equilibrium.
On general grounds, it could be expected that $\tau_{relax}$ were 
proportional to the mean free collision
time $\tau_{coll}$.

We define the relaxation time for a quadrupolar deformation 
as \cite{26}:
$${ 1 \over \tau_{relax}} = 
{{\int ({\partial{n}\over{\partial{t}}})_{col} 
~P_2(\theta)~ d{\vec p}} \over 
{\int n(\vec p})~ P_2(\theta)~ d{\vec p}}
\eqno(10)$$
To evaluate eq.(10) numerically, 
we deform the initial equilibrium distribution as:
$${\vec p} \to p~ (1 + \alpha P_2(\theta)) 
\eqno(11)$$ 
where $P_2 (\theta)= {1 \over 2} [3 cos^2 (\theta) - 1]$ is  
the Legendre function. We note that for an arbitrary monopolar 
deformation (that implies only a scaling change in the modulus of the momenta 
$p \to p (1 + \alpha)$) the Boltzmann-Nordheim
collision integral in eq.(1) becomes exactly zero 
(this holds also for fermions and
classical particles). In this case the dynamics is entirely governed 
by the external (mean-field) potential. 

The quadrupolar relaxation time as a function of temperature is shown in Fig.(3)
(solid line) for $\alpha = 0.5$. 
We plot in the
same figure the mean free 
collision time $\tau_{coll}= {dN_{coll} \over {dt}})^{-1}$,
corresponding to the average time between two
collisions for the system in the thermal equilibrium. 
These two characteristic times are proportional 
for $T / T_c > 3$, 
but their behavior differs significantly at lower
temperatures. For $ T/T_c < 2.7$
the collision time drops due to the increase in the collision rate (see fig.1)
but, surprisingly enough,
the relaxation time increases as:
$$\tau_{rel} \propto {(T - T_c)}^{-1/2} \eqno(2)$$
Note that for a classical
Maxwell-Boltzmann distribution the relaxation time goes as 
$\tau_{relax}^{MB} \propto {T}^{-1/2}$. 
The reason for this behavior can be understood looking at the
expression of the collision integral, eq.(1). 
The two terms involving the product of four distribution functions,
$F_4 =  n(\vec r, \vec p) n(\vec r, \vec p_2)
n(\vec r, \vec p_3) n(\vec r, \vec p_4)$ 
cancel each other exactly for any distribution $n(\vec r, \vec p, t)$. 
On the other hand, 
it is this term that, in eq.(5), gives the largest contribution 
to the collision rate at low temperatures. 
Moreover, the terms involving
the product of three distributions,
$F_3 =  n(\vec r, \vec p) n(\vec r, \vec p_2)
[ n(\vec r, \vec p_3) +  n(\vec r, \vec p_4)]$,
almost cancel each other for the deformation that
we have considered.
In fig.(4,a) 
we show separatedly the contributions of each term of eq.(1) 
to the relaxation of the
initial quadrupolar deformation.
The dot-dashed line indicates the time evolution of the quadrupolar
momentum considering only
the term $F_4$ in the Monte Carlo simulation:
as expected there 
is no contribution to the relaxation of the system. However, in fig.(4,b),
we show how this term gives the largest contribution to the 
collision rate.
The term $F_3$ gives also a small contribution in fig.(4,a) (dotted line),
while it still gives a large contribution to the collision rate in fig.(4,b). 
Hence, we end up with the conclusion that
the stimulated collisions
do not contribute to the relaxation of the system.
Only the 
spontaneous collision
term ($F = 1$ in eq.(4))  
affects the relaxation time,
although this gives the smallest contribution to the collision rate 
(long-dashed line in fig.s(4,a,b)).

As a consequence of these results, we remark that the simple
criteria given in the
literature that state the transition between zero and first sound occurs
when $\omega~\tau_{coll} << 1$, with $\omega$ the frequency of a collective
mode,
do not apply for a boson gas at 
$T_c < T < 2.7~ T_c$, where $\tau_{coll} \to 0$, but $\tau_{rel} \to \infty$. 
Close to the critical temperature, the system should exhibit collisionless
modes, with a transition to first sound at higher temperatures, when 
$\omega \tau_{relax} <<1$.

The predicted trend of $\tau_{relax}$ vs. temperature 
in fig.(3) can be 
tested experimentally.
In particular, the damping of collective
quadrupole motions, induced in the trapped gas,
should rapidly decrease approaching $T_c$ 
according eq.(12). In effect the B-E distribution observed
experimentally is non-homogeneous, being the boson gas trapped in an
harmonic field. However we do expect that the conclusion of this work and 
the trend predicted by eq.(12)
can still give a good approximation provided that the size
of the system is much larger than the characteristic scale of inhomogeneity.

We conclude remarking that the rapid increase
of $\tau_{relax}$ for $T \to T_c$
questions the assumption that the condensation
observed experimentally takes place in equilibrated systems.
Understanding how the BEC occurs in systems far from
equilibrium is one of the most interesting and challenging
problems raised by the physics of trapped atoms. 

\vskip 0.8truecm

Part of this work has been done at the 
the Physics Department of the University
of
Illinois at Urbana-Champaign.
T.L-A. thanks D.K.Campbell and 
A.S. thanks V.R.Pandharipande
for the kind hospitality
and support.

Discussions with G.Baym, S. Fantoni, G.M.Kavoulakis, S.R.Shenoy and
J.P.Wolfe are acknowledged.

\pagebreak

\pagebreak

\begin{figure}
\caption{ \noindent
Normalized collision rate  
as a function
of the temperature, calculated numerically (solid line)
and analitycally (dotted line). The dashed line shows the 
collision rate for a classical 
Maxwell-Boltzmann distribution.
The particle density is $\rho = {10}^{13} {cm}^{-3}$ and
the collision cross section is $\sigma = 4.4~{10}^{-13} {cm}^{2} $.
}
\end{figure}

\begin{figure}
\caption{ \noindent
Contributions to the collision rate 
from each term of the collision integral in BNE eq.(1), as explained
in the text.
}
\end{figure}

\begin{figure}
\caption{ \noindent 
Comparison between the relaxation time (solid line) 
and the mean free collision time 
(dashed line)
as a function of temperature.
}
\end{figure}

\begin{figure}
\caption{ \noindent
(a) Quadrupole moment as a function of time for $T = 100~nK$. 
The different lines correspond to contributions coming from
each term in the
collision integral of BEN eq.(1), as explained in the text (see fig.2).
(b) Corresponding collision rates.}
\end{figure}



\begin{references}
\bibitem{1} M.H.Anderson, J.R.Ensher, M.R.Matthews, C.E.Wieman and
E.A.Cornell, Science 269 (1995) 198
\bibitem{2}K.B.Davis et al.,Phys.Rev.Lett.75, (1995) 3969
\bibitem{3}C.C.Bradley et al., Phys.Rev.Lett.75 (1995) 1687; 77 (1997) 985
\bibitem{4}G.M.Kavoulakis, G.Baym and J.P.Wolfe, Phys.Rev.B53 (1996)7227
and ref.s therein
\bibitem{5}For a review see
"Bose-Einstein condensation"
edited by A.Griffin, D.W.Snoke, and S.Stringari
(Cambridge University Press, Cambridge, England, 1995).
\bibitem{6} 
Yu. Kagan contribution in \cite{5}.
\bibitem{7} H.T.C. Stoof contribution in [5]. 
\bibitem{8} Gardiner and Zoller, Phys.Rev. A55 (1997) 2902
\bibitem{9}G.Baym and C.J.Pethick, Phys.Rev.Lett.76 (1996) 6
\bibitem{10} V.V.Goldman, I.Silvera and A.Legget, Phys. Rev. B24 (1981) 2870
\bibitem{11} D.A.Huse and ED. Siggia, Jour. of Low Temp. Phys. 46 (1982) 137
\bibitem{12} C. E. Wieman, Am. J. Phys. 64 (1996) 847 
\bibitem{13} E. Levich and V. Yakho, J. Phys. A11 (1978) 2237
\bibitem{14}D.W.Snoke and J.P.Wolfe, Phys.Rev.B39 (1989) 4030
\bibitem{15}D.V.Semikoz and I.I.Tkachev, Phys.Rev.Lett.74 (1995) 3093
\bibitem{16} D.S. Jin et al., Phys.Rev.Lett.77 (1996) 420
\bibitem{17} M.O.Meves et al. Phys.Rev.Lett. 77 (1996) 988
\bibitem{18} Yu.Kagan. et al. Phys.Rev.A54 (1996) 1753R
\bibitem{19} A. Smerzi and S. Fantoni, Phys. Rev. Lett. 78 (1997) 3589
\bibitem{20} S. Stringari, Phys. Rev. Lett. 77 (1996) 2360; 
M. Edwards et al. Phys. Rev. Lett. 77 (1996) 1671
\bibitem{21} D.S.Jin et al., Phys. Rev. Lett. 78 (1997) 764
\bibitem{22}C.W.Nordheim, Proc.R.Soc. London, A119 (1928) 689
\bibitem{23}G.F.Bertsch and S.Gupta, Phys.Rep.160 (1988) 189
\bibitem{24}
 A.Bonasera, F.Gulminelli and J.Molitoris, Phys.Rep.243, (1994) 1
\bibitem{25}G.M.Welke and G.F.Bertsch, Phys.Rev.C45 (1992) 1403;
G. F. Bertsch in "Statistical description of transport in plasma, 
astro- and nuclear physics", edited by J.Misguich, G. Pelletier and
P. Schuck, Les Houches, February 1993. 
\bibitem{26} G.Bertsch, Z.Physik A (1978) 103
\bibitem{27} A.Griffin, Wen-Chin Wu and S. Stringari Phys. Rev. Lett.
78 (1997) 1838 and references therein

\end{references}
\end{document}